\documentclass[12pt]{article}
\usepackage[
	colorlinks=true,
	citecolor=black,
	linkcolor=black,
	urlcolor=blue,
	hypertexnames=false]{hyperref}

\usepackage[normalem]{ulem}

\def\bean{\begin{equation*}}
\def\eean{\end{equation*}}
\newcommand{\arXiv}[2]{\href{http://arxiv.org/pdf/#1}{{\tt #2/#1}}}
\newcommand{\arXivold}[1]{\href{http://arxiv.org/pdf/#1}{{\tt #1}}}

\def\be{\begin{equation}}
\def\ee{\end{equation}}
\def\ba{\begin{eqnarray}}
\def\ea{\end{eqnarray}}

\def\bdm{\begin{displaymath}}
\def\edm{\end{displaymath}}
\def\la{~\mbox{\raisebox{-.6ex}{$\stackrel{<}{\sim}$}}~}
\def\ga{~\mbox{\raisebox{-.6ex}{$\stackrel{>}{\sim}$}}~}
\def\bq{\begin{quote}}
\def\eq{\end{quote}}

 at 10truept

\newcommand{\beq}{\begin{equation}}
\newcommand{\eeq}{\end{equation}}
\newcommand{\bea}{\begin{eqnarray}}
\newcommand{\eea}{\end{eqnarray}}
\newcommand{\beqa}{\begin{eqnarray}}
\newcommand{\eeqa}{\end{eqnarray}}

\def\la{~\mbox{\raisebox{-.6ex}{$\stackrel{<}{\sim}$}}~}
\def\ga{~\mbox{\raisebox{-.6ex}{$\stackrel{>}{\sim}$}}~}

\def\ltap{\ \raise.3ex\hbox{$<$\kern-.75em\lower1ex\hbox{$\sim$}}\ }
\def\gtap{\ \raise.3ex\hbox{$>$\kern-.75em\lower1ex\hbox{$\sim$}}\ }
\def\gl{\ \raise.5ex\hbox{$>$}\kern-.8em\lower.5ex\hbox{$<$}\ }
\def\roughly#1{\raise.3ex\hbox{$#1$\kern-.75em\lower1ex\hbox{$\sim$}}}

\begin{document}

\thispagestyle{empty}
\begin{flushright}
October 2017
\end{flushright}
\vspace*{1.35cm}
\begin{center}
{\Large \bf Landscaping the Strong $CP$ Problem}\\

\vspace*{1.25cm} {\large Nemanja Kaloper\footnote{\tt
kaloper@physics.ucdavis.edu} and John Terning\footnote{\tt
jterning@gmail.com}}\\
\vspace{.5cm} {\em Department of Physics, University of
California, Davis, CA 95616, USA}\\

\vspace{1.5cm} 
\end{center}

\begin{abstract}
\noindent 
One often hears that the strong $CP$ problem is {\em the} one problem which cannot be solved by anthropic reasoning. We argue that this is not so. Due to nonperturbative dynamics, states with a different $CP$ violating paramenter $\theta$ acquire different vacuum energies after the QCD phase transition. These add to the total variation of the cosmological constant in the putative landscape of Universes. 
An interesting possibility arises when the cosmological constant is mostly cancelled by the membrane nucleation mechanism. 
If the step size in the resulting discretuum of cosmological constants, $\Delta \Lambda$, is in the interval $({\rm meV})^4 < \Delta \Lambda <  (100 \, {\rm MeV})^4$, the cancellation of vacuum energy can be assisted by the scanning of $\theta$. 
For $({\rm meV})^4 < \Delta \Lambda < ({\rm keV})^4$ this yields $\theta < 10^{-10}$, meeting the 
observational limits. This scenario opens up 24 orders of magnitude of acceptable parameter space for $\Delta \Lambda$ compared to membrane nucleation acting alone. 
In such a Universe one may not need a light axion to solve the strong $CP$ problem.
\end{abstract}

\vfill \setcounter{page}{0} \setcounter{footnote}{0}
\newpage

The Anthropic Principle has been touted as a possible explanation for a variety of curious physical facts about our Universe that make it consistent with our existence. The most famous case is that of the cosmological constant. The anthropic explanation of the smallness of the cosmological constant \cite{davies,lindeanthropics,banks,Weinberg:1987dv} is the only widely accepted solution to this particular mystery. The difficulties\footnote{A proposal to use global dynamics to stabilize the cosmological constant has been made recently in \cite{Kaloper:2013zca,Kaloper:2015jra}.} in finding alternative explanations have led many physicists who previously scoffed at the Anthropic Principle to embrace it. Occasionally some of the curious facts, like the ratio of the weak scale to the strong scale, turn out to have little to do with  anthropic explanations since our existence could be ensured with completely different laws of physics \cite{Harnik:2006vj}. It still seems worthwhile, however,  to use anthropic reasoning if only to further understand its explanatory power and limitations.
There are, of course, many variants of the Anthropic Principle, ranging from the egotistical---``the Universe is set up to produce intelligent life like us"---to the tautological---``the Universe must be just as it is, otherwise it would be different." We will restrict ourselves to the weak form, as used by Weinberg \cite{Weinberg:1987dv}, which merely requires self-consistency.

The strong $CP$ problem refers to another curious fact concerning our Universe. For completeness we will briefly remind the reader of how the problem crops up. The nontrivial gauge group topology in non-Abelian gauge theories gives rise to a complex vacuum structure
which can be represented as a smooth manifold parameterized by a phase $\theta_0$ varying  continuously in the
interval $[0,2\pi]$. In the full Lagrangian, this phase appears as the coefficient of the topological term 
\beq
Q= \frac{g^2 }{16 \pi^2} {\rm Tr} \,G_{\mu\nu} \, {}^*G^{\mu\nu}~.
\eeq
 In the presence of fermions charged under the gauge group, the axial current is not conserved due to the ABJ anomaly \cite{Adler:1969gk}. Hence chiral transformations mix $\theta_0$ with the  overall phase of the fermion mass matrix, yielding 
an effective angle $\theta$ \cite{tHooft:1979rat}
\beq
\theta = \theta_0+ {\rm Arg}\, {\rm det } M ~.
\eeq
If $\det M = 0$, its phase is completely arbitrary and can be changed at will. In this case $\theta_0$ is completely arbitrary and unphysical as well---and it can be set to zero without any effect on the rest of the theory. 

In our Universe it seems that all quarks are massive, with masses arising from Yukawa couplings to the Higgs field, which are generically complex. With $\det M \ne 0$, 
there is no way to cancel the phase associated with the strong interactions. Absorbing $\theta_0$ into the quark mass matrix by a chiral transformation yields a $CP$-violating neutron-pion coupling, set by $\theta$ which controls the value of the neutron dipole moment \cite{Crewther:1979pi}. 
The limits on the neutron dipole moment imply that $\theta$ is bounded by \cite{Pospelov:1999ha,Guo:2015tla} 
\beq
\theta < 10^{-10} \, .
\label{thetabound}
\eeq
We are left with the problem of understanding why the otherwise arbitrary value of $\theta$ is so small (modulo $2\pi$). Alternatively, 
the question is why do the otherwise arbitrary values of $\theta_0$ and ${\rm Arg} \, {\rm det} M$ cancel with a precision of at least $\la 10^{-10}$. 

A commonly invoked answer is Peccei-Quinn (PQ) $U(1)$ symmetry breaking \cite{Peccei:1977hh}, 
resulting in the Weinberg-Wilczek axion \cite{Weinberg:1977ma,Wilczek:1977pj}, which is the Goldstone boson of the broken PQ symmetry. In this approach, $\theta$ is the vacuum expectation value of a field which has minima naturally very close to $2 n \pi$. 
A small value of $\theta$ might be a consequence of symmetry breaking at high scales, with the 
value of $\theta$ set by irrelevant operators \cite{Nelson:1983zb,Barr:1984qx}. 

Having the 
neutron dipole moment much larger than the observational bound (\ref{thetabound}) has little effect on the real world \cite{Banks:2003es,Donoghue:2003vs,Donoghue:2007zz}. In particular the important processes of cosmogenesis appear to be completely blind to it, suggesting that the value of $\theta$ is essentially irrelevant, affecting nothing but 
the largely peripheral neutron dipole moment. 
Thus it would seem that the
gross irrelevance of  the neutron dipole moment precludes any chance for resorting to an anthropic argument to explain the smallness of $\theta$. 

In this Letter we shall argue otherwise. 
Instantons generate $\theta$-dependent vacuum energy contributions, and gravity then gives different $\theta$ vacua different cosmological histories. These contributions to the vacuum energy only materialize after the QCD phase transition in the later Universe, but since their scale is $\sim (100 \, {\rm MeV})^4 \gg ({\rm meV})^4$, they should also be cancelled by whatever neutralizes the energy of the vacuum. If anthropics is the answer to the cosmological constant problem \cite{Weinberg:1987dv,Bousso:2000xa}, it should also account for the QCD contributions to the vacuum energy.

We will frame our argument  in the setting of the anthropic landscape  of string theory \cite{Bousso:2000xa}. In this context the vacuum energy is neutralized using membranes charged under 3-form fields, whose field  strength flux contributions to vacuum energy can be discharged by membrane emission \cite{Brown:1987dd}. 
These leaps in the value of the cosmological constant must meet certain requirements for an anthropic argument to work.

Firstly, to avoid simply fine-tuning the final value to the observed one, the leaps should allow for a discretuum\footnote{For an example of a very dense discretuum with $\Delta \Lambda \ll ({\rm meV})^4$ see \cite{Banks:1991mb}.} of possible vacuum energies with a spacing $\Delta \Lambda \simeq \Lambda_{\rm observed} \sim ({\rm meV})^4$. This can happen if the theory includes a large number of  form fields, with many possible values of $\Lambda$ that differ by $\Delta \Lambda$. This can be arranged by some crafty model building \cite{Bousso:2000xa}. Assume that the total effective cosmological constant is
\beq
\Lambda = \frac12 \sum^J_i n_i^2 q_i^2 + \Lambda_{\rm regularized} \, ,
\eeq
where the first term is the contribution from $4$-form fluxes $F_{(i)} = n_i q_i$ where $n_i$ is the number of units of the membrane charge $q_i$, and the second term accounts for vacuum energy contributions calculated from all other degrees of freedom in that particular vacuum.  It must be assumed that $\Lambda_{\rm regularized}$ is negative.
Since $\Lambda_{\rm regularized} \propto - (\Lambda_{UV})^4$, where $\Lambda_{UV}$ is the UV cutoff,
the flux contributions should be $\sum_i n_i^2 q_i^2  \ga 2 |\Lambda_{\rm regularized}|$, such that initially 
$\Lambda \sim \Lambda_{UV}^4 \gg 0$. Such initial states are typical. 

To cancel $\Lambda_{\rm regularized}$ to a given precision $\Delta \Lambda$, one needs flux states which satisfy 
\beq
2 |\Lambda_{\rm regularized}| < \sum_i n_i^2 q_i^2 < 2 (|\Lambda_{\rm regularized}| + \Delta \Lambda)~.
\eeq 
This is the equation for a spherical shell in $J$ dimensions, of volume 
\beq
{\cal V} \simeq 
\omega_{J-1} (2| \Lambda_{\rm regularized}|)^{J/2-1} \Delta \Lambda~,
\eeq
where $\omega_{J-1}$ is the volume
of the $J$-dimensional sphere of unit radius. The shell will contain at least one flux configuration if its volume is greater than the unit cell volume $D\Pi_i q_i$, where $D$ counts degeneracy of the states, which can be quite large. So the spacing between nearby states in the cosmological constant discretuum is \cite{Bousso:2000xa} 
\beq
\Delta \Lambda = \frac{D \Pi_i q_i}{\omega_{J-1} (2|\Lambda_{\rm regularized}|)^{J/2-1}} \, .
\label{lmin}
\eeq
If this is true for $\Lambda_{\rm regularized} \sim - \Lambda_{UV}^4$ calculated to some order in perturbation theory, it will remain true order by order in the loop expansion. All one then needs is to model-build the theory (a.k.a., compactify the relevant higher dimensional supergravity on a manifold that yields the right low energy theory to reproduce the Standard Model, and supports a system of forms and membranes yielding (\ref{lmin}))
to achieve the required precision $\Delta \Lambda$. 

The initial state is a highly curved de Sitter vacuum. 
Inside large $\Lambda$ regions, membranes are nucleated leading to a 
cascade of bubbles inside which the cosmological constant is reduced. 
As $\Lambda$ drops, the membrane nucleation rate slows down. 
Also, the transitions involving multi-membrane emissions, simultaneously discharging many units of flux can 
be suppressed in some regions of the landscape.
This ensures that inflation is not impeded by  very rapid discharge of vacuum energy.
Finally gravity suppresses the transitions to states with large negative cosmological constant \cite{Coleman:1980aw}. This means that the states with small $\Lambda$, positive or negative, will be metastable.

As we noted, the leaps should be slow since otherwise they would discharge the vacuum energy too fast, and prevent inflation from ever taking place. On the other hand if the leaps are too slow, they could continue well past the inflaton has slow-rolled to its minimum. If that happened, the Universe would have continued to inflate for far too long, without any significant reheating taking place, ending up empty and devoid of structures \cite{Abbott:1984qf}. This would nullify any benefit from an interim stage of slow roll inflation.

The ``empty Universe'' problem can be avoided in regions where the initial cosmological constant 
$\Lambda$ overwhelms the inflationary potential \cite{Bousso:2000xa}. As long as this is true after
the penultimate jump, the Universe in the penultimate bubble will be undergoing eternal inflation, with a random distribution of inflaton values. When the ultimate jump happens inside this region, in the interior of the bubble the cosmological constant will sharply drop. Eternal inflation will terminate, and slow roll inflation can occur
yielding reheating and seeding curvature perturbations in the final Universe with a small final $\Lambda$.

If $\Delta \Lambda \simeq ({\rm meV})^4 \simeq 10^{-120} (M_{Pl})^4$, one 
can invoke Weinberg's anthropic argument \cite{Weinberg:1987dv} and its refinements
\cite{Garriga:1999bf} to pick the terminal value of the cosmological constant. Basically, if $\Delta \Lambda \simeq ({\rm meV})^4$, then one naturally favors the values of $- \Delta \Lambda < \Lambda < \Delta \Lambda$. With an additional assumption of their uniform 
apriori 
distribution, one finds that the favored value is $\Lambda \simeq \Delta \Lambda$, fitting observation. 

What if $\Delta \Lambda > ({\rm meV})^4$? In the regions of the landscape where this occurs, small terminal values of $\Lambda$ close to the observed value would not seem to be typical. One might therefore conclude that
anthropic reasoning would not help in this case since such regions would be uninhabitable. Yet such corners will occur in the landscape for various reasons: too few form fields, very large degeneracies, wrong values of charges, and so on. Ignoring the question of which regions are more typical (we don't know), we wish to simply point out that dismissing such regions is premature. In fact, many regions of the landscape where $\Delta \Lambda$ induced by membrane charges is larger than $({\rm meV})^4$ may allow for a different, perhaps even more curious, anthropic approach to the cosmological constant problem. Concretely, if $({\rm mev})^4 < \Delta \Lambda < ({\rm keV})^4$, anthropic reasoning combined with membrane nucleation dynamics and some assumptions about inflation and
reheating may yield a simultaneous solution of both the cosmological constant problem {\it and} the strong $CP$ problem!
  
Enter QCD. The lifting of the degeneracy between the QCD $\theta$-vacua occurs {\it after} the QCD phase transition due to the strong coupling phenomena which generate a potential $V_{\rm QCD}(\theta)$ for $\theta$. 
The potential adds to the cosmological constant, but only {\it after} the QCD phase transition. In general 
$V_{\rm QCD}(\theta)$ is a periodic function of $\theta$.  As Vafa and Witten have shown using a general path integral argument \cite{Vafa:1984xg}, 
the minima of $V_{\rm QCD}$ occur at $\theta=0, {\rm mod }\,2 \pi$, which are the only $CP$ invariant vacua. 

If we expand $V_{\rm QCD}(\theta)$ in a Taylor series 
around the vacuum $\theta = 0$, where $V_{\rm QCD}'(\theta)|_{\theta=0} = V_1 = 0$, we obtain
\beq
V_{\rm QCD}(\theta) = \sum_n \frac{V_n}{n!} \theta^n = V_0 +\frac{1}{2}V_2\, \theta^2+\ldots \, .
\eeq
The coefficient of the second order term is related to the topological susceptibility of QCD (up to an equal time commutator) \cite{Luscher:1978rn,Witten:1979vv,Urban:2009vy}, 
\beq
\frac{d^2V_{\rm QCD}}{d\theta^2} = \int d^4x \langle0| T Q(x) Q(0) |0 \rangle~.
\eeq
 One can estimate it using the large $N$ limit \cite{Witten:1979vv}, where this term is 
 \beq
 V_2 = \frac{m_{\eta^\prime}^2 f_\pi^2}{6} \sim  10 \, (100 \, {\rm MeV})^4~.
 \eeq
Lattice simulations \cite{Petreczky:2016vrs,Borsanyi:2016ksw} currently give $V_2 \approx (75 \,{\rm MeV})^4$,
we will simply parameterize it as 
\beq
V_2 =  (a \,100 \, {\rm MeV})^4~,
\eeq
where $a\sim {\mathcal O}(1)$.

After the final membrane nucleation the effective value of $\Lambda$ is small enough so that 
the inflaton potential can dominate and ordinary inflation begins. After inflation ends, and the Universe reheats, there are still other phase transitions that are yet to occur.  Certainly for our scenario to work the QCD phase transition is still to occur and also possibly (depending on the reheat temperature) the electroweak phase transition and (more hypothetically) even a GUT phase transition.
Now, if $\Delta \Lambda \sim ({\rm meV})^4$, for any value of $\theta$ and $V_0$ (as well as contributions from other phase transitions \cite{Linde:1978px,Bellazzini:2015wva}) there will still be many flux states with a final value of $\Lambda$ which differ
from each other by $\Delta \Lambda$.  Therefore in these regions of the landscape the arguments of  \cite{Bousso:2000xa} remain unaffected, and one cannot find any useful conclusions about the value of $\theta$ using anthropic reasoning. Simply put,
the scanning of $\theta$ is screened by membrane emission.

On the other hand, suppose that $\Delta \Lambda > ({\rm meV})^4$. When this happens, the flux scanning induced by membrane emission cannot naturally yield states with $\Lambda \sim ({\rm meV})^4$. In the absence of any other free parameters that  can scan a range of values, one might infer that anthropic reasoning alone would not be sufficient to pick $\Lambda$ in Weinberg's window \cite{Weinberg:1987dv}
\beq
-({\rm meV})^4 < \Lambda <(3\, {\rm meV})^4~,
\label{eq:window}
\eeq
as noted in \cite{Bousso:2000xa}. 

Yet in our case there is the value of $\theta$ which can be scanned  over continuously \cite{andrei}. Since we are interested in states inside Weinberg's window at very late times (i.e. now!), we can combine the flux scanning with 
large steps $\Delta \Lambda$ with scanning in $\theta$. The idea is that flux scanning brings the cosmological constant as close as possible to Weinberg's window, and $\theta$ scanning does the rest in order for the overall final value
of $\Lambda$
to meet the anthropic requirements. Since the $\frac{1}{2} V_2 \theta^2$ correction is positive, this means that we need
$\frac{1}{2} V_2 \theta^2 - \Delta \Lambda$ to be comparable to the cosmological constant now, $\sim ({\rm meV})^4$. This means that the Anthropic Principle favors the values of $\theta$ that mostly cancel the larger contributions from $\Delta \Lambda$ down to $({\rm meV})^4$, in a way which is completely analogous to using the Anthropic Principle to pick the counterterm that cancels the regulated value of the vacuum energy in \cite{Weinberg:1987dv}. Analogous to \cite{Bousso:2000xa} we need $\Lambda$ after flux scanning added to $V_0$ to be negative so that the $\theta$ dependent term can cancel it. This does not affect vacuum stability after the final membrane emission since $V_0 \sim -(100\, {\rm MeV})^4$, so the net 
cosmological energy density can be positive at that time.

For $({\rm keV})^4  < \Delta \Lambda < (100 {\rm  \, MeV})^4$ successful scanning requires $\theta > 10^{-10}$. While the cosmological constant can be reduced to the observed value, the required $\theta$ is too large. This rules out this class of solutions, which demonstrates that our suggestion is experimentally falsifiable.  

However with $({\rm meV})^4  < \Delta \Lambda < ({\rm keV})^4$ the cancellation of 
$\Delta \Lambda$ which allows the current $\Lambda$ to saturate the anthropic bound 
requires $\theta  <10^{-10}$. Yet while the spacing of the discretuum $\Delta \Lambda $ is small, one might worry that
the magnitude of $\Lambda$ after the QCD phase transition that needs to be canceled by $\theta$ scanning can be much larger than $\Delta \Lambda$. For example, with $\Delta \Lambda = ({\rm keV})^4$, $|\Lambda|$ could be as large as $(100\, {\rm MeV})^4$ and still be cancelled by $\theta$ scanning.
For this to happen $\theta$  must be more finely scanned for larger values of $|\Lambda|$ in order to meet the 
anthropic requirements. Indeed, for the final value of the cosmological constant to be in Weinberg's window (\ref{eq:window}), $\theta$ must be scanned to reach $\sqrt{2\Lambda/V_2}$ with an accuracy of $\sim ({\rm meV})^4/\sqrt{\Lambda\,V_2}$. If the distribution of $\theta$ is approximately uniform, 
the most probable region to find $\theta$ in is where $\Lambda \sim \Delta \Lambda$. In this region the values 
of $\theta$ (that saturate the bound) depend on $\Delta \Lambda$, and range in the interval
\beq
10^{-22} < \theta  < 10^{-10} \, .
\eeq

We have found the most probable value of $\theta$, but this still leaves a very large number of other less probable Universes where $|\Lambda| \gg \Delta \Lambda$ can be cancelled by larger values of $\theta$. So what is the expected value of $\theta$? This brings in two of the perennial pests of landscaping: the problem of probability distributions and the question of what observables should be held fixed. Given a full theory one would hope to resolve the latter question, but the former problem might never be resolved.

Let us first consider what happens if we hold both the inflaton sector and the baryon to photon ratio, $n_B/n_\gamma$, fixed. Fixing the inflaton sector ensures that the reheat temperature and the density perturbations at the end of inflation (i.e. the seeds of galaxies) are held constant; if the density perturbations can change then Weinberg's window also changes, since their size determines how long it takes structures to form. Fixing
$n_B/n_\gamma \sim 10^{-9}$ ensures that we get the usual Big Bang Nucleosynthesis.
Since nothing depends on $\theta$ before the QCD phase transition, Universes with different values of $\theta$
share almost exactly the same early cosmology. 
When inflation ends reheating occurrs
at some temperature $T_*$, producing the usual visible contents of the Universe and also the dark matter sector.
Just after reheating, almost all of the energy density is stored in the form of relativistic particles, a.k.a. radiation. As the Universe cools, some particle energy densities begin to scale like matter, with the protons and neutrons arising at the QCD 
phase transition, $T_{QCD}\sim$ 100 MeV. 
The dark matter sector starts to behave like cold dark matter at a temperature scale $M$, where $M<T_*$.

Now, let us compare 
two Universes, say with $\theta \ne 0$ and $\theta=0$. In the Universe with
$\theta \ne 0$ we see that below the QCD scale part of the energy density is taken up by the  vacuum energy contribution $V_2 \theta^2$. That is, $V_2 \theta^2$ acts like a latent heat, so after the QCD phase transition the Universe with larger $\theta$ has a lower temperature at the same scale factor. Since the
total comoving energy is fixed by cosmology at the end of inflation, much before the QCD phase transition, the amount of dark matter in the Universe with $\theta \ne 0$ must be smaller than in the Universe with $\theta = 0$, for all else being equal\footnote{This is similar to Linde's argument for
axionic dark matter \cite{Linde:1991km}.}. To estimate the required decrease of dark matter,  
\beq
\delta \rho_{DM}(T,\theta) = \rho_{DM}(T,\theta) - \rho_{DM}(T,0)~,  
\eeq
we can scale the amount of extra visible radiation required ($\rho_R=V_2\theta^2$) from $T_{QCD}$ up to  the temperature $M$. This yields 
\beq
\delta \rho_{DM}(T_{QCD},\theta) \simeq  -V_2 \theta^2 \frac{M}{T_{QCD}}~.
\eeq
Since the baryon comoving energy density is fixed below $T_{QCD}$, the ratio of dark matter to baryon energy
density is constant from here on, and so we find that
\be
\frac{\delta \rho_{DM}}{\rho_{baryon}} \simeq - \frac{V_2 \theta^2}{10^{-9}\, T_{QCD}^4} \frac{M}{T_{QCD}} \, .
\ee
Thus in the Universe with $\theta \ne 0$ the onset of matter domination is delayed relative to a Universe with $\theta = 0$. As a result  galaxies 
will begin to form later, but this cannot be delayed too long \cite{Hellerman:2005yi} since it must occur before the cosmological constant dominates; this is just Weinberg's original anthropic requirement. Restricting 
$\delta \rho_{DM}/\rho_{baryon}$ to within an order of magnitude and using $V_2/T_{QCD}^4 \sim 1$, we obtain
\be
\theta^2 \le 10^{-8} \,\frac{T_{QCD}}{M} \, .
\ee
If $M\sim {\rm TeV}$, this imposes the bound $\theta < 10^{-6}$.
In such cases, to get down to the most probable band of $\Lambda \sim \Delta \Lambda$ we'd still have to tune,
but only by one part in $10^4$. On the other hand if $M \ga 10^8 \, {\rm TeV}$, the bound yields 
$\theta < 10^{-10}$, consistent with the observed limit without any fine-tuning.

This argument shows that the expected value of $\theta$ is model dependent, and it can be suppressed 
by further anthropic requirements. Other factors may also suppress the expected value of $\theta$.
Just to showcase this, recall that the geometric probability picture which suggests that large values of $\theta$ are expected is based
on the width of the allowed values of $\theta$ in each band, yielding a probability  $p_\theta \sim 1/\sqrt{\Lambda V_2}$ that is proportional to the scanning 
accuracy. Here $\Lambda = -n \Delta \Lambda$ is discrete, running between $-V_2$ and $-\Delta \Lambda$, so
$p_\theta \sim 1/\sqrt{n}$. Since for a fixed $n$ the value of $\theta$ which places the net cosmological constant in 
Weinberg's window is $V_2 \theta^2  - \Lambda \la {\rm meV}^4$, we see that $\theta \sim \sqrt{\Lambda} \sim \sqrt{n}$. The expected value over the whole ensemble of bands is $\sum_n \theta \,p_\theta / \sum_n p_\theta $ when all values of $\Lambda$ are equally probable. This distribution without any further cutoffs would seem to favor $\theta \sim {\cal O}(1)$. Clearly additional anthropic restrictions, such as the one induced by requiring a sufficient amount of dark matter, can reduce it dramatically. Further, any additional suppression of large
vales of  $\Lambda$ would also suppress the expected value of $\theta$. In general there is some (unknown) probability distribution, $p_\Lambda$, for different values of $\Lambda$. A falling probability distribution does not solve the cosmological constant problem by itself if $\Delta \Lambda \gg ({\rm meV})^4$. This means that the scanning in $\theta$ is still needed. Hence the probability to find a value of $\theta$ which sufficiently cancels $\Lambda$ becomes $p_\theta \, p_\Lambda$. So the expectation value of $\theta$ will depend on the unknown distribution $p_\Lambda$ in addition to further anthropic selection criteria like having the right abundance of dark matter. 

Hence we see that there may be some boroughs of the landscape where anthropic reasoning can explain both the observed smallness of the cosmological constant and the CP violating QCD angle $\theta$. Note, that allowing for the variation of $\theta$ relaxes the constraint on $\Delta \Lambda$ needed to cancel the vacuum energy down to the observed value by as much as $24$ orders of magnitude. Also note that if experimental accuracy in measuring the neutron dipole moment  eventually improves to the point where $\theta <  10^{-22}$, then again we could rule out our scenario. To be sure, these considerations are very model dependent. Yet, it does seem plausible that a vast landscape may contain such regions. Our considerations therefore---in the very least---show that we cannot exclude the possibility that anthropic reasoning lies behind both the smallness of the cosmological constant and the QCD $\theta$ parameter.

While it is always difficult for an anthropic argument to avoid the whiff of a ``Just So Story,"
there is an experimental consequence. In such a Universe solving the strong $CP$ problem does not require a light axion. Since there is much work underway dedicated to looking for the QCD axion we might find out soon whether it exists---or not. The absence of the QCD axion might point us in the anthropic direction. If so, 
the anthropically favored value of $\theta$ should approximately saturate the bound. Hence accurately measuring the neutron dipole moment, and in turn $\theta$, could yield estimates of $\Delta \Lambda$ (larger than the size of Weinberg's window) that could be compared to more detailed landscape scenarios. 

In our view however the most important implications are conceptual. Up until now it has been widely thought that the strong $CP$ problem is not prone to anthropic solutions \cite{Banks:2003es,Donoghue:2003vs}, aside from the case of irrational axion \cite{Banks:1991mb}. There are in fact arguments that anthropic reasoning supports the QCD axion as the {\it natural} solution of the strong $CP$ problem \cite{Arvanitaki:2009fg}. Now this is not so clear. We hope that the arguments presented here will at least stimulate discussion that could shed more light on this question, which has been raised only extremely rarely so far \cite{Banks:1991mb,Weiss:1987ns,Takahashi:2009zzd}.

We also can't help but wonder, since life in our Universe is often ironic \cite{Harnik:2006vj,Morissette}, what further ironies lie ahead? Given that an anthropic explanation of the strong $CP$ problem took so long to identify even though it simply links the strong $CP$ and cosmological constant problems\footnote{Which have been likened in the past \cite{Aurilia:1980xj,Wilczek:1983as}.}, could there be, as yet unidentified, neighborhoods of the multiverse where a non-anthropic, dynamical relaxation of the cosmological constant is in effect?

\section*{Acknowledgements}

We thank H.-C. Cheng, C. Cheung, G. D'Amico, R. Kitano, M. Kleban, A. Lawrence, A. Linde, and P. Saraswat for interesting discussions and comments. This work was supported in part by DOE grant DE-SC0009999.

\end{document}